\documentclass[reprint, amsmath, amssymb, aps]{revtex4-2}
\usepackage{orcidlink,graphicx,cmap}
\usepackage[utf8]{inputenc}
\usepackage[T1]{fontenc}

\def\re#1{\mathrm{Re}(#1)}
\def\im#1{\mathrm{Im}(#1)}
\def\K{{\cal K}}
\def\Order#1{{\cal O}\left(#1\right)}

\begin{document}
\title{Quasinormal ringing of a regular black hole sourced by the Dehnen-type distribution of matter}
\author{S. V. Bolokhov \orcidlink{0000-0002-9533-530X}}\email{bolokhov-sv@rudn.ru}
\affiliation{RUDN University, 6 Miklukho-Maklaya St, Moscow, 117198, Russian Federation}
\begin{abstract}
We study quasinormal modes of test scalar, electromagnetic, and Dirac fields in the background of a new analytic regular black-hole solution obtained as an exact solution of the Einstein equations sourced by a Dehnen-type matter distribution in [R. A. Konoplya, A. Zhidenko, arXiv:2511.03066]. The metric is asymptotically flat and characterized by a simple lapse function
$
f(r)=1-2 M r^{2}/(r+a)^{3},
$
where $M$ is the ADM mass and $a$ represents the characteristic scale of the surrounding dark-matter halo that regularizes the central region. The effective potentials for all perturbing fields possess the standard single-barrier form, ensuring linear stability and the applicability of the WKB formalism. The quasinormal frequencies are computed using the sixth- and ninth-order WKB methods with Pad\'e corrections and verified by time-domain integration, both approaches showing excellent agreement. The parameter $a$ leads to a moderate increase in the real oscillation frequency, while the damping rate remains almost unchanged, producing only small corrections to the Schwarzschild spectrum. Since such deviations become appreciable only for unrealistically dense halos, our results confirm that the quasinormal spectrum of astrophysical black holes is safely unaffected by ordinary galactic dark-matter environments.
\end{abstract}

\maketitle

\section{Introduction}

Quasinormal modes (QNMs) represent the characteristic oscillations of black holes and other compact objects under small perturbations. Their complex frequencies encode both the oscillation rate and the damping time of perturbations, forming a unique fingerprint of the underlying spacetime geometry. In the era of gravitational-wave astronomy, quasinormal spectra have become one of the most powerful tools for testing general relativity and probing potential deviations arising from quantum corrections, modified gravity, or environmental effects~\cite{Kokkotas:1999bd,Berti:2009kk,Konoplya:2011qq,Cardoso:2016rao,Bolokhov:2025uxz}. Along with QNMs, grey-body factors---which quantify the partial reflection and transmission of radiation through the effective potential barrier---are essential for understanding black-hole scattering and Hawking radiation spectra.

Recently, a new class of \emph{regular} and asymptotically flat black-hole solutions has been constructed by coupling Einstein gravity to an anisotropic fluid mimicking a galactic dark-matter halo~\cite{Konoplya:2025ect}. In contrast to many earlier approaches, this construction in the case of the Dehnen-type distribution of matter produces remarkably simple exact analytic metrics that are everywhere regular, asymptotically flat. The resulting spacetimes interpolate smoothly between a Schwarzschild-like exterior and a de Sitter-like core, without invoking exotic matter or violating the weak energy condition. They therefore provide realistic and self-consistent models of supermassive black holes embedded in galactic environments.

The dynamical response of such objects to external perturbations is a natural and necessary step toward understanding their stability and observational signatures. In the present work, we compute the quasinormal spectra of scalar, electromagnetic, and Dirac test fields in the background of these regular halo-supported black holes. The analysis is performed using the sixth-order WKB approximation with Pad\'e resummation, which offers high accuracy for low overtones and moderate multipole numbers~\cite{Konoplya:2019hlu,Matyjasek:2017psv}. Selected results are further verified by direct time-domain integration of the perturbation equations, providing an independent consistency check.

By comparing the spectra for different field spins and parameters of the dark-matter halo, we aim to clarify how the presence of the halo modifies the oscillation frequencies, damping rates, and scattering properties of the regular black holes. The obtained results may also serve as a reference point for future studies of gravitational perturbations, grey-body factors, and possible observational imprints of such configurations.

The paper is organized as follows. 
In Sec.~\ref{sec:wavelike} we introduce the Konoplya--Zhidenko metric, obtained as an exact solution of the Einstein equations with a Dehnen-type dark-matter distribution, and derive the master wave equations governing scalar, electromagnetic, and Dirac perturbations in this background. 
Sec.~\ref{sec:WKB} describes the higher-order WKB formalism with Pad\'e approximants, which is used to compute the dominant quasinormal frequencies, while Sec.~\ref{sec:time} outlines the time-domain integration technique employed to verify the results using an independent numerical approach. 
The quasinormal spectra obtained for various spin fields and multipole numbers are analyzed in Sec.~\ref{sec:QNM}, where we also discuss the convergence of the WKB expansion and the agreement with the time-domain data. 
Sec.~\ref{sec:QNM} examines also the eikonal limit and its connection with the parameters of null geodesics, allowing one to relate the high-frequency oscillations to the shadow radius and Lyapunov exponent. 
The correspondence between quasinormal modes and grey-body factors is presented as well. 
Finally, the main findings and physical implications of the study are summarized in Sec.~\ref{sec:Conclusions}.

\section{Black hole metric and effective potentials}\label{sec:wavelike}


The spacetime considered here corresponds to the regular, asymptotically flat black hole obtained in~\cite{Konoplya:2025ect}, where Einstein gravity is coupled to an anisotropic fluid representing the dark-matter halo. The geometry is described by the static and spherically symmetric line element
\begin{equation}\label{metric}
ds^{2}=-f(r)\,dt^{2}+\frac{dr^{2}}{f(r)}+r^{2}\left(d\theta^{2}+\sin^{2}\theta\,d\phi^{2}\right),
\end{equation}
where the metric function reads
\begin{equation}\label{fmetric}
f(r)=1-\frac{2Mr^{2}}{(r+a)^{3}}.
\end{equation}
Here $M$ is the ADM mass of the black hole, and $a>0$ is a parameter encoding the characteristic scale of the surrounding dark-matter distribution. The simple metric above is a solution of the Einstein equation with the matter distribution \cite{Dehnen:1993uh,Taylor:2002zd}
\begin{equation}\label{Dehnendensity}
\rho(r)=\rho_0 \left(\frac{r}{a}\right)^{-\alpha} \left(1+\frac{r^k}{a^k}\right)^{-(\gamma-\alpha)/k}
\end{equation}
with $\alpha=0$, $\gamma=4$, and $k=1$.
For $r\gg a$ the function approaches the Schwarzschild form $f(r)\simeq 1-2M/r$, while near the origin it behaves as $f(r)\simeq 1-2r^{2}/a^{2}$, indicating a regular de~Sitter core.  
Throughout this work we adopt geometrized units with $G=c=1$ and set $M=1$, so that all quantities are measured in units of the black-hole mass.


To probe the dynamical response of this spacetime, we consider minimally coupled test fields of spin $s=0$, $1/2$, and $1$, corresponding to scalar, Dirac, and electromagnetic perturbations, respectively.  
The covariant field equations can be written in the unified form
\begin{subequations}\label{coveqs}
\begin{align}
{(-g)}^{-1/2}\,\partial_{\mu}\!\left(\sqrt{-g} \,g^{\mu\nu}\,\partial_{\nu}\Phi\right)&=0, \label{KGg}\\
{(-g)}^{-1/2}\,\partial_{\mu}\!\left(F_{\rho\sigma}g^{\rho\nu}g^{\sigma\mu}\sqrt{-g}\right)&=0,\label{EmagEq}\\
\gamma^{\alpha}\!\left(\frac{\partial}{\partial x^{\alpha}}-\Gamma_{\alpha}\right)\Upsilon&=0,\label{covdirac}
\end{align}
\end{subequations}
where $\Phi$ is the scalar field, $A_{\mu}$ the electromagnetic four-potential with $F_{\mu\nu}=\partial_{\mu}A_{\nu}-\partial_{\nu}A_{\mu}$, and $\Upsilon$ the Dirac spinor.  
The matrices $\gamma^{\alpha}$ are defined in a locally orthonormal tetrad basis, and $\Gamma_{\alpha}$ denote the corresponding spin connections.


Owing to the spherical symmetry of the background, all perturbation equations (\ref{coveqs}) admit separation of variables.  
For the scalar field, one can use the ansatz (see, for instance \cite{Konoplya:2018arm})
\[
\Phi(t,r,\theta,\phi)=\frac{\Psi(r)}{r}\,Y_{\ell m}(\theta,\phi)\,e^{-i\omega t},
\]
where $Y_{\ell m}(\theta,\phi)$ are the spherical harmonics.  
A similar decomposition applies to the electromagnetic and Dirac fields using spin-weighted harmonics \cite{Konoplya:2011qq}.  
After separating the angular and temporal variables, all perturbations reduce to a Schrödinger-type wave equation of the form
\begin{equation}\label{wave-equation}
\frac{d^{2}\Psi}{dr_{*}^{2}}+\!\left[\omega^{2}-V(r)\right]\!\Psi=0,
\end{equation}
where $r_{*}$ is the tortoise coordinate defined by
\begin{equation}\label{tortoise}
\frac{dr_{*}}{dr}=\frac{1}{f(r)}.
\end{equation}


For integer-spin fields, the effective potential can be expressed compactly as
\begin{equation}\label{potentialScalar}
V(r)=f(r)\left[\frac{\ell(\ell+1)}{r^{2}}+(1-s)\frac{1}{r}\frac{df(r)}{dr}\right],
\end{equation}
where $\ell=s,s+1,\dots$ is the multipole number and $s=0,1$ corresponds to the scalar and electromagnetic cases, respectively.

For the Dirac field ($s=1/2$), one obtains a pair of supersymmetric partner potentials~\cite{Chandrasekhar:1976ap},
\begin{equation}
V_{\pm}(r)=W^{2}\pm\frac{dW}{dr_{*}},\qquad
W=\left(\ell+\frac{1}{2}\right)\frac{\sqrt{f(r)}}{r}.
\end{equation}
The two potentials $V_{+}$ and $V_{-}$ are isospectral; their wave functions are connected by the Darboux transformation,
\begin{equation}\label{psi}
\Psi_{+}\propto\left(W+\frac{d}{dr_{*}}\right)\Psi_{-}.
\end{equation}
Consequently, it is sufficient to compute the quasinormal frequencies for only one of them, and we choose $V_{+}(r)$ since the WKB method demonstrates better numerical convergence in this case.

In what follows, we will employ these effective potentials to study the characteristic quasinormal spectra and grey-body factors for scalar, electromagnetic, and Dirac test fields in the background of the regular Konoplya--Zhidenko black hole.

\begin{figure}
\resizebox{\linewidth}{!}{\includegraphics{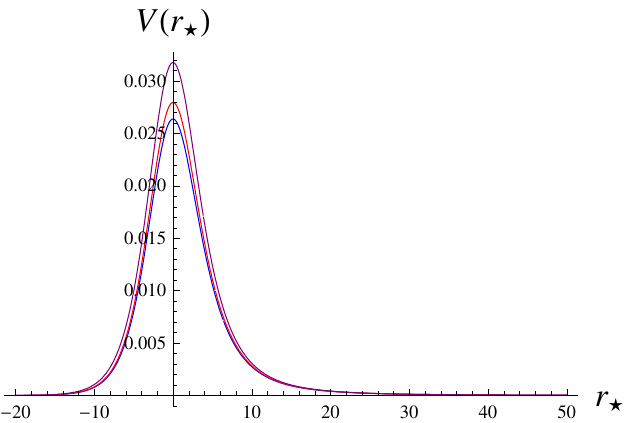}}
\caption{Effective potential as a function of the tortoise coordinate $r_{*}$ for $\ell=0$ scalar field perturbations: $M=1$; $a=0$ (blue), $a=0.05$ (red) and $a=0.15$ (purple).}\label{fig:scalarpotL0}
\end{figure}

\begin{figure}
\resizebox{\linewidth}{!}{\includegraphics{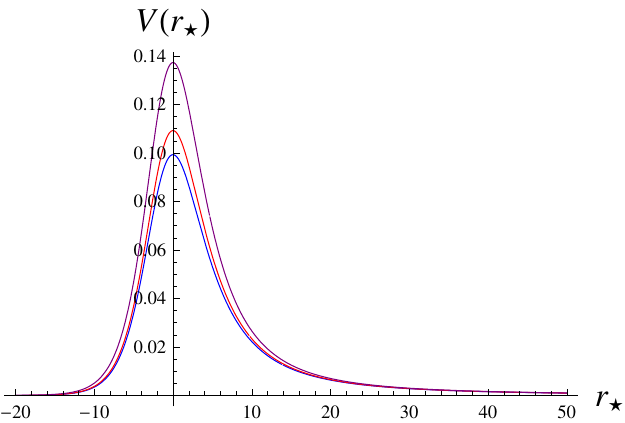}}
\caption{Effective potential as a function of the tortoise coordinate $r_{*}$ for $\ell=1$ scalar field perturbations: $M=1$; $a=0$ (blue), $a=0.05$ (red) and $a=0.15$ (purple).}\label{fig:scalarpotL1}
\end{figure}

\begin{figure}
\resizebox{\linewidth}{!}{\includegraphics{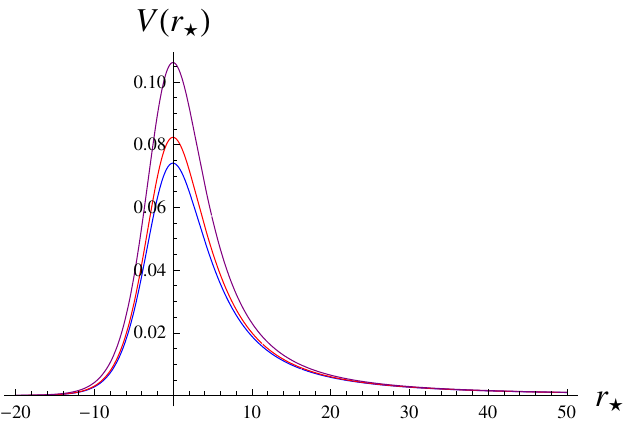}}
\caption{Effective potential as a function of the tortoise coordinate $r_{*}$ for $\ell=1$ electromagnetic field perturbations: $M=1$; $a=0$ (blue), $a=0.05$ (red) and $a=0.15$ (purple).}\label{fig:EMpot}
\end{figure}

\begin{figure}
\resizebox{\linewidth}{!}{\includegraphics{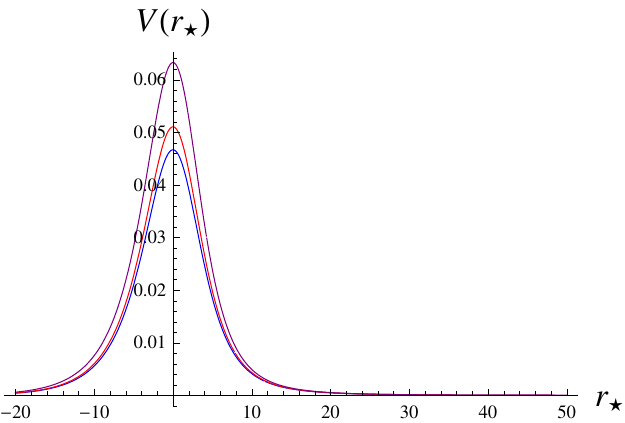}}
\caption{Effective potential as a function of the tortoise coordinate $r_{*}$ for $\ell=1/2$ Dirac field perturbations: $M=1$; $a=0$ (blue), $a=0.05$ (red) and $a=0.15$ (purple).}\label{fig:Diracpot}
\end{figure}

\section{WKB formalism for quasinormal modes}\label{sec:WKB}


When the effective potential $V(r)$ appearing in the master equation~(\ref{wave-equation}) forms a single-peaked potential barrier that tends to constant asymptotic values on both sides, the quasinormal frequencies can be obtained using the semi-analytic Wentzel–Kramers–Brillouin (WKB) approximation.  
The relevant boundary conditions correspond to purely ingoing waves at the event horizon and purely outgoing waves at spatial infinity,
\begin{equation}\label{boundaryconditions}
\Psi(r_* \to -\infty) \propto e^{-i\omega r_*}, \qquad
\Psi(r_* \to +\infty) \propto e^{+i\omega r_*}.
\end{equation}
These conditions select a discrete set of complex frequencies $\omega=\omega_{\text{R}}-i\omega_{\text{I}}$ describing the damped oscillations characteristic of black-hole perturbations.

The WKB method provides an efficient way to approximate such spectra when the potential barrier is smooth and possesses a single maximum, as in the case of most test fields around static spherically symmetric black holes \cite{Skvortsova:2023zmj,Lutfuoglu:2025ljm,Bolokhov:2025egl,Zhao:2022gxl,del-Corral:2022kbk,Lutfuoglu:2025pzi,Skvortsova:2024atk,Bolokhov:2022rqv,Skvortsova:2024wly,Kodama:2009bf,Lutfuoglu:2025blw}. The approach consists in matching the WKB expansions of the wave function in three regions—on both sides of and near the peak of the potential—and enforcing the continuity of the logarithmic derivatives at the turning points.  
The lowest (first) WKB order corresponds to the eikonal limit and becomes exact for $\ell\to\infty$, where the wave oscillates much faster than the potential varies.


The higher-order WKB scheme generalizes this procedure by systematically expanding the complex frequency in powers of the small parameter $\mathcal{K}^{-1}$, where
\begin{equation}
\mathcal{K}=n+\frac{1}{2}, \qquad n=0,1,2,\ldots
\end{equation}
is the overtone number.   In the vicinity of the potential maximum $r=r_{0}$, one introduces the derivatives
\[
V_{i} \equiv \left.\frac{d^{i}V}{dr_{*}^{i}}\right|_{r_{*}(r_{0})}, 
\]
and expresses the squared frequency as an asymptotic expansion around the eikonal limit~\cite{Iyer:1986np,Konoplya:2003ii,Matyjasek:2017psv},
\begin{align}\label{WKBformula-spherical}
\omega^{2} &= V_{0}
  + A_{2}(\mathcal{K}^{2})
  + A_{4}(\mathcal{K}^{2})
  + A_{6}(\mathcal{K}^{2})
  + \ldots \nonumber \\
&\quad - i\,\mathcal{K}\sqrt{-2V_{2}}
  \Bigl[1 + A_{3}(\mathcal{K}^{2}) + A_{5}(\mathcal{K}^{2}) + \ldots \Bigr],
\end{align}
where $V_{0}$ and $V_{2}$ are, respectively, the potential and its second derivative at the maximum.  
The coefficients $A_{i}$ denote successive WKB corrections depending on $\mathcal{K}$ and the higher derivatives of $V(r)$ at $r=r_{0}$ up to order $2i$.  
Explicit expressions are known up to the thirteenth order: $A_{2}$ and $A_{3}$ were derived in~\cite{Iyer:1986np},  
$A_{4}$–$A_{6}$ in~\cite{Konoplya:2003ii}, and  
$A_{7}$–$A_{13}$ in~\cite{Matyjasek:2017psv}.

The WKB expansion~(\ref{WKBformula-spherical}) rapidly converges for the fundamental and first few overtones ($n < \ell$), provided the potential barrier is smooth and the field propagation is of short wavelength compared to the curvature scale (see \cite{Skvortsova:2024eqi, Bonanno:2025dry, Lutfuoglu:2025eik, Lutfuoglu:2025qkt, Konoplya:2023ahd, Bolokhov:2024ixe, Bolokhov:2023bwm, Bolokhov:2023ruj, Arbelaez:2025gwj} for recent examples).
However, higher orders do not always yield a monotonic improvement in precision.  
To stabilize convergence and improve accuracy, it is advantageous to resum the series using Pad\'e approximants constructed from the WKB expansion of $\omega^{2}$~\cite{Konoplya:2019hlu}.  
The Pad\'e–WKB method for the low-laying modes typically achieves an accuracy nearly equivalent to the most precise numerical techniques such as the continued-fraction (Leaver) method.

This approach has been widely applied in the literature for evaluating quasinormal modes and grey-body factors across a broad class of black-hole backgrounds and field spins ~\cite{Abdalla:2005hu,Zhidenko:2003wq,Konoplya:2001ji,Konoplya:2022hbl,Kokkotas:2010zd,
Ishihara:2008re,Albuquerque:2023lhm,Paul:2023eep,Guo:2022hjp}.
In the present work we use the sixth-order WKB formula improved by Pad\'e approximants as our main computational tool for determining the quasinormal spectra of scalar, electromagnetic, and Dirac fields in the Konoplya--Zhidenko regular black-hole background.

\section{Time-domain integration method}\label{sec:time}


An independent verification of the quasinormal frequencies obtained by the WKB approach can be achieved through the direct numerical integration of the perturbation equations in the time domain.  
This method captures the entire dynamical evolution of a perturbation, including the initial transient, the quasinormal ringing phase, and the late-time tails.  
It is based on integrating the master equation~(\ref{wave-equation}) in the characteristic (light-cone) coordinates,
\begin{equation}
u = t - r_{*}, \qquad v = t + r_{*},
\end{equation}
where $r_{*}$ is the tortoise coordinate defined by Eq.~(\ref{tortoise}).  
In these coordinates, the wave equation takes the form
\begin{equation}\label{wave-uv}
4\,\frac{\partial^{2}\Psi}{\partial u\,\partial v}
+ V(r)\,\Psi = 0.
\end{equation}


To evolve this equation numerically, we employ the well-known Gundlach–Price–Pullin finite-difference scheme~\cite{Gundlach:1993tp}, which is second-order convergent and stable for smooth potentials.  
The computational grid is constructed by discretizing the $(u,v)$ plane with a constant step size $\Delta$, so that the function $\Psi$ is evaluated at the four corners of each elementary null square:  
$N=(u+\Delta,v+\Delta)$, $W=(u+\Delta,v)$, $E=(u,v+\Delta)$, and $S=(u,v)$.  
The field value at the new point $N$ is obtained iteratively as
\begin{equation}\label{gppscheme}
\Psi_{N} = \Psi_{W} + \Psi_{E} - \Psi_{S}
- \frac{\Delta^{2}}{8}\,V\!\left(r_{S}\right)
\left(\Psi_{W} + \Psi_{E}\right)
+ \mathcal{O}(\Delta^{4}),
\end{equation}
where $r_{S}$ corresponds to the radial coordinate along the diagonal of the cell.  
The evolution begins with initial data specified on two null surfaces $u=u_{0}$ and $v=v_{0}$, typically chosen as a compact Gaussian pulse localized near the potential peak,
\begin{equation}
\Psi(u=u_{0},v) = \exp\!\left[-\frac{(v-v_{c})^{2}}{2\sigma^{2}}\right], \qquad
\Psi(u,v=v_{0}) = 1,
\end{equation}
where $v_{c}=v_{0}$ determines the center of the pulse and $\sigma$ its width.  
This choice avoids numerical contamination from the artificial initial burst and efficiently excites the fundamental and low-overtone modes.


Once the field evolution is computed, the signal $\Psi(t,r)$ at a fixed radial position is analyzed to extract the complex frequencies of the exponentially damped oscillations during the ringdown stage.  
The relevant part of the waveform can be fitted by a superposition of quasinormal modes,
\begin{equation}
\Psi(t,r)\approx
\sum_{j} C_{j}\, e^{-i\omega_{j} t},
\end{equation}
where $\omega_{j}=\omega_{\mathrm{R},j}-i\omega_{\mathrm{I},j}$ and $C_{j}$ are complex amplitudes.  
The fitting is performed using the Prony method, which reconstructs the exponential parameters directly from the discrete time series of the numerically integrated signal.  
This technique allows one to determine the dominant modes with high precision and serves as an effective cross-check for the WKB--Pad\'e results.


The time-domain integration method has the advantage of being independent of any assumption about the shape of the potential barrier (see examples in \cite{Cuyubamba:2016cug,Konoplya:2025uiq,Konoplya:2013sba,Skvortsova:2023zca,Konoplya:2024lch,Malik:2024tuf,Qian:2022kaq,Momennia:2022tug,
Aneesh:2018hlp,Malik:2024nhy,Malik:2024elk,Konoplya:2007jv,Dubinsky:2024aeu,Dubinsky:2024mwd,Bronnikov:2021liv,Malik:2024qsz,Dubinsky:2024gwo}) and is thus well suited for cases in which the potential exhibits multiple peaks or when the WKB expansion converges poorly.  Moreover, it provides valuable information on the late-time tails, which are inaccessible to frequency-domain techniques.

However, when the field mass $\mu$ is sufficiently large, the late-time regime becomes dominated by oscillatory tails, and the exponential quasinormal ringing phase shortens considerably.  
In such cases, extracting the complex frequencies through the Prony fit becomes challenging, and the accuracy of the determined modes deteriorates.  
For the massless fields studied here, the method performs exceptionally well, and the extracted frequencies agree with the WKB--Pad\'e results to within a fraction of one percent for the fundamental mode.

In the next section we use both the WKB--Pad\'e approximation and the time-domain analysis described above to compute the quasinormal spectra of scalar, electromagnetic, and Dirac perturbations in the Konoplya--Zhidenko regular black-hole background.

\begin{figure}
\resizebox{\linewidth}{!}{\includegraphics{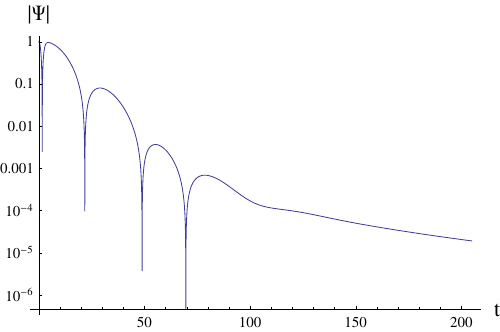}}
\caption{Time-domain profile for the scalar perturbations at $\ell=0$ and $a=0.1$, $M=1$. The WKB data for the fundamental frequency is $\omega = 0.124786 - 0.106453 i$ and the time-domain method gives $\omega = 0.123736 - 0.106996 i$. The difference of about one percent should be attributed to the error of the Prony method, because the period of quasinormal ringing is very short for $\ell=0$ case. }\label{fig:TDscalarL0}
\end{figure}

\begin{figure}
\resizebox{\linewidth}{!}{\includegraphics{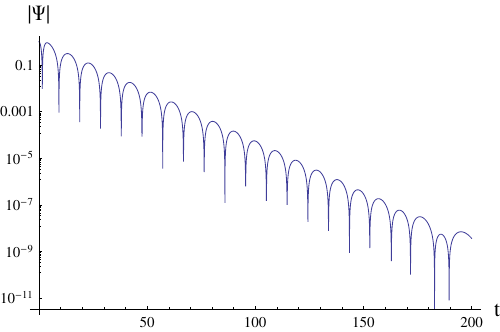}}
\caption{Time-domain profile for the scalar perturbations at $\ell=1$ and $a=0.1$, $M=1$. The WKB data for the fundamental frequency is $\omega = 0.327921 - 0.100512 i$ and the time-domain method gives $\omega = 0.327912 - 0.100507 i$. The difference is about $\sim 10^{-6}$. }\label{fig:TDscalarL1}
\end{figure}

\begin{figure}
\resizebox{\linewidth}{!}{\includegraphics{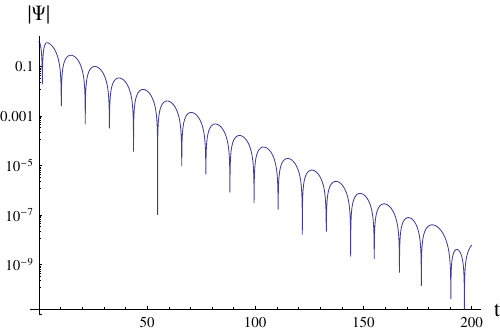}}
\caption{Time-domain profile for the Maxwell perturbations at $\ell=1$ and $a=0.1$, $M=1$. The WKB data for the fundamental frequency is $\omega = 0.282020 - 0.096084 i$ and the time-domain method gives $\omega = 0.282033 - 0.0960883 i$. The difference is about $\sim 10^{-5}$. }\label{fig:TDMaxwellL1}
\end{figure}

\begin{figure}
\resizebox{\linewidth}{!}{\includegraphics{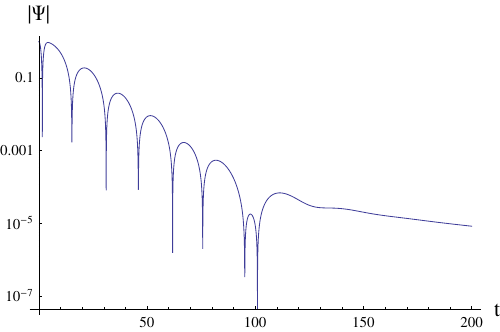}}
\caption{Time-domain profile for the Dirac perturbations at $\ell=1/2$ and $a=0.1$, $M=1$. The WKB data for the fundamental frequency is $\omega = 0.206108 - 0.099664 i$ and the time-domain method gives $\omega = 0.206552 - 0.0998381 i$. The difference is a small fraction of one percent.}\label{fig:TDDirac}
\end{figure}

\begin{table}[ht]
\centering
\small
\begin{tabular}{c c c c}
\hline
\hline
$a$ & WKB$_6$ ($m=3$) & WKB$_9$ ($m=4$) & \!\!\!\!\!\!Difference [\%] \\
\hline
0.00 & $0.111950 - 0.104580i$ & $0.111444 - 0.104504i$ & 0.334 \\
0.02 & $0.114253 - 0.105026i$ & $0.113750 - 0.104976i$ & 0.325 \\
0.04 & $0.116678 - 0.105445i$ & $0.116179 - 0.105424i$ & 0.318 \\
0.06 & $0.119235 - 0.105830i$ & $0.118743 - 0.105838i$ & 0.309 \\
0.08 & $0.121934 - 0.106170i$ & $0.121457 - 0.106210i$ & 0.296 \\
0.10 & $0.124786 - 0.106453i$ & $0.124337 - 0.106525i$ & 0.277 \\
0.12 & $0.127807 - 0.106664i$ & $0.127403 - 0.106765i$ & 0.250 \\
0.14 & $0.131015 - 0.106778i$ & $0.130673 - 0.106900i$ & 0.215 \\
0.16 & $0.134448 - 0.106759i$ & $0.134168 - 0.106891i$ & 0.180 \\
0.18 & $0.138184 - 0.106552i$ & $0.137900 - 0.106668i$ & 0.176 \\
0.20 & $0.142343 - 0.105992i$ & $0.141818 - 0.106017i$ & 0.296 \\
\hline
\hline
\end{tabular}
\caption{Fundamental quasinormal modes of the scalar field with $\ell=0$ for the Konoplya--Zhidenko black hole ($M=1$), computed using the sixth- and ninth-order WKB methods with Pad\'e approximants. The relative deviation between the two approximations is given in percent.}
\end{table}

\begin{table}[ht]
\centering
\small
\begin{tabular}{c c c c}
\hline
\hline
$a$ & WKB$_6$ ($m=3$) & WKB$_9$ ($m=4$) & \!\!\!\!\!\!Difference [\%] \\
\hline
0.00 & $0.292931 - 0.097660i$ & $0.292935 - 0.097659i$ & 0.0013 \\
0.02 & $0.299048 - 0.098271i$ & $0.299049 - 0.098269i$ & 0.0008 \\
0.04 & $0.305552 - 0.098869i$ & $0.305551 - 0.098866i$ & 0.0010 \\
0.06 & $0.312493 - 0.099449i$ & $0.312489 - 0.099445i$ & 0.0016 \\
0.08 & $0.319926 - 0.100000i$ & $0.319919 - 0.099996i$ & 0.0024 \\
0.10 & $0.327921 - 0.100512i$ & $0.327910 - 0.100508i$ & 0.0033 \\
0.12 & $0.336562 - 0.100967i$ & $0.336547 - 0.100964i$ & 0.0042 \\
0.14 & $0.345955 - 0.101341i$ & $0.345936 - 0.101340i$ & 0.0053 \\
0.16 & $0.356235 - 0.101600i$ & $0.356211 - 0.101603i$ & 0.0063 \\
0.18 & $0.367574 - 0.101693i$ & $0.367547 - 0.101703i$ & 0.0074 \\
0.20 & $0.380204 - 0.101542i$ & $0.380177 - 0.101561i$ & 0.0084 \\
\hline
\hline
\end{tabular}
\caption{Quasinormal frequencies of the scalar field with $\ell=1$ for the Konoplya--Zhidenko black hole ($M=1$). The results from the sixth- and ninth-order WKB approximations with Pad\'e corrections show excellent convergence, with differences below $0.01\%$.}
\end{table}

\begin{table}[ht]
\centering
\small
\begin{tabular}{c c c c}
\hline
\hline
$a$ & WKB$_6$ ($m=3$) & WKB$_9$ ($m=4$) & \!\!\!\!\!\!Difference [\%] \\
\hline
0.00 & $0.248251 - 0.092479i$ & $0.248264 - 0.092488i$ & 0.0058 \\
0.02 & $0.254104 - 0.093223i$ & $0.254118 - 0.093231i$ & 0.0058 \\
0.04 & $0.260352 - 0.093962i$ & $0.260365 - 0.093970i$ & 0.0054 \\
0.06 & $0.267044 - 0.094691i$ & $0.267056 - 0.094698i$ & 0.0050 \\
0.08 & $0.274242 - 0.095402i$ & $0.274254 - 0.095408i$ & 0.0045 \\
0.10 & $0.282020 - 0.096084i$ & $0.282031 - 0.096089i$ & 0.0040 \\
0.12 & $0.290469 - 0.096720i$ & $0.290479 - 0.096724i$ & 0.0035 \\
0.14 & $0.299705 - 0.097288i$ & $0.299714 - 0.097291i$ & 0.0029 \\
0.16 & $0.309875 - 0.097755i$ & $0.309881 - 0.097754i$ & 0.0019 \\
0.18 & $0.321175 - 0.098070i$ & $0.321177 - 0.098068i$ & 0.0009 \\
0.20 & $0.333869 - 0.098152i$ & $0.333865 - 0.098148i$ & 0.0017 \\
\hline
\hline
\end{tabular}
\caption{Quasinormal frequencies of the scalar field with $\ell=1$ obtained via the sixth- and ninth-order WKB approximations. The results demonstrate nearly perfect agreement, confirming the robustness of the WKB--Pad\'e method.}
\end{table}

\begin{table}[ht]
\centering
\small
\begin{tabular}{c c c c}
\hline
\hline
$a$ & WKB$_6$ ($m=3$) & WKB$_9$ ($m=4$) & \!\!\!\!\!\!Difference [\%] \\
\hline
0.00 & $0.457595 - 0.095005i$ & $0.457596 - 0.095004i$ & 0.00015 \\
0.02 & $0.467450 - 0.095672i$ & $0.467451 - 0.095671i$ & 0.00015 \\
0.04 & $0.477956 - 0.096331i$ & $0.477956 - 0.096331i$ & 0.00014 \\
0.06 & $0.489193 - 0.096978i$ & $0.489194 - 0.096978i$ & 0.00014 \\
0.08 & $0.501261 - 0.097602i$ & $0.501262 - 0.097602i$ & 0.00013 \\
0.10 & $0.514281 - 0.098194i$ & $0.514282 - 0.098193i$ & 0.00011 \\
0.12 & $0.528402 - 0.098735i$ & $0.528403 - 0.098735i$ & 0.00010 \\
0.14 & $0.543811 - 0.099205i$ & $0.543811 - 0.099205i$ & 0.00008 \\
0.16 & $0.560748 - 0.099570i$ & $0.560748 - 0.099569i$ & 0.00004 \\
0.18 & $0.579529 - 0.099778i$ & $0.579529 - 0.099778i$ & 0.00002 \\
0.20 & $0.600586 - 0.099751i$ & $0.600586 - 0.099751i$ & 0.00005 \\
\hline
\hline
\end{tabular}
\caption{Fundamental quasinormal modes of the scalar field with $\ell=2$ for the Konoplya--Zhidenko black hole ($M=1$). The sixth- and ninth-order WKB--Pad\'e approximations agree up to $10^{-4}\%$.}
\end{table}

\begin{table}[ht]
\centering
\small
\begin{tabular}{c c c c}
\hline
\hline
$a$ & WKB$_6$ ($m=3$) & WKB$_9$ ($m=4$) & \!\!\!\!\!\!Difference [\%] \\
\hline
0.00 & $0.182638 - 0.096583i$ & $0.182745 - 0.097026i$ & 0.221 \\
0.02 & $0.186705 - 0.097221i$ & $0.186858 - 0.097638i$ & 0.211 \\
0.04 & $0.191046 - 0.097857i$ & $0.191243 - 0.098239i$ & 0.200 \\
0.06 & $0.195696 - 0.098484i$ & $0.195933 - 0.098823i$ & 0.189 \\
0.08 & $0.200700 - 0.099091i$ & $0.200969 - 0.099381i$ & 0.177 \\
0.10 & $0.206108 - 0.099664i$ & $0.206401 - 0.099900i$ & 0.164 \\
0.12 & $0.211983 - 0.100186i$ & $0.212290 - 0.100364i$ & 0.151 \\
0.14 & $0.218404 - 0.100628i$ & $0.218713 - 0.100749i$ & 0.138 \\
0.16 & $0.225467 - 0.100954i$ & $0.225768 - 0.101019i$ & 0.125 \\
0.18 & $0.233299 - 0.101107i$ & $0.233583 - 0.101122i$ & 0.112 \\
0.20 & $0.242070 - 0.101003i$ & $0.242328 - 0.100975i$ & 0.0988 \\
\hline
\hline
\end{tabular}
\caption{Quasinormal frequencies of the Dirac field with $\ell = 1/2$ for the Konoplya--Zhidenko black hole ($M=1$). Results from the sixth- and ninth-order WKB--Pad\'e approximations show subpercent-level deviations that decrease with increasing $a$.}
\end{table}

\begin{table}[ht]
\centering
\small
\begin{tabular}{c c c c}
\hline
\hline
$a$ & WKB$_6$ ($m=3$) & WKB$_9$ ($m=4$) & \!\!\!\!\!\!Difference [\%] \\
\hline
0.00 & $0.380034 - 0.096400i$ & $0.380037 - 0.096405i$ & 0.0015 \\
0.02 & $0.388041 - 0.097034i$ & $0.388044 - 0.097038i$ & 0.0013 \\
0.04 & $0.396572 - 0.097658i$ & $0.396575 - 0.097662i$ & 0.0012 \\
0.06 & $0.405692 - 0.098266i$ & $0.405695 - 0.098270i$ & 0.0011 \\
0.08 & $0.415482 - 0.098850i$ & $0.415486 - 0.098853i$ & 0.00099 \\
0.10 & $0.426038 - 0.099398i$ & $0.426041 - 0.099401i$ & 0.00093 \\
0.12 & $0.437479 - 0.099894i$ & $0.437482 - 0.099896i$ & 0.00089 \\
0.14 & $0.449954 - 0.100314i$ & $0.449957 - 0.100316i$ & 0.00086 \\
0.16 & $0.463655 - 0.100624i$ & $0.463658 - 0.100626i$ & 0.00084 \\
0.18 & $0.478833 - 0.100773i$ & $0.478836 - 0.100776i$ & 0.00084 \\
0.20 & $0.495829 - 0.100683i$ & $0.495833 - 0.100685i$ & 0.00085 \\
\hline
\hline
\end{tabular}
\caption{Quasinormal frequencies of the Dirac field with $\ell = 3/2$ for the Konoplya--Zhidenko black hole ($M=1$). Both WKB--Pad\'e approximations yield practically identical values, confirming convergence at the level of $10^{-3}\%$.}
\end{table}

\begin{table}[ht]
\centering
\small
\begin{tabular}{c c c c}
\hline
\hline
$a$ & WKB$_6$ ($m=3$) & WKB$_9$ ($m=4$) & \!\!\!\!\!\!Difference [\%] \\
\hline
0.00 & $0.436532 - 0.290726i$ & $0.436544 - 0.290725i$ & 0.0021 \\
0.02 & $0.446886 - 0.292586i$ & $0.446893 - 0.292577i$ & 0.0021 \\
0.04 & $0.457926 - 0.294412i$ & $0.457931 - 0.294398i$ & 0.0028 \\
0.06 & $0.469737 - 0.296184i$ & $0.469740 - 0.296166i$ & 0.0033 \\
0.08 & $0.482424 - 0.297874i$ & $0.482426 - 0.297854i$ & 0.0036 \\
0.10 & $0.496113 - 0.299447i$ & $0.496114 - 0.299426i$ & 0.0036 \\
0.12 & $0.510959 - 0.300852i$ & $0.510959 - 0.300832i$ & 0.0034 \\
0.14 & $0.527157 - 0.302017i$ & $0.527156 - 0.301998i$ & 0.0030 \\
0.16 & $0.544954 - 0.302837i$ & $0.544952 - 0.302821i$ & 0.0025 \\
0.18 & $0.564675 - 0.303154i$ & $0.564672 - 0.303143i$ & 0.0018 \\
0.20 & $0.586752 - 0.302724i$ & $0.586748 - 0.302719i$ & 0.0010 \\
\hline
\hline
\end{tabular}
\caption{First overtone ($n=1$) quasinormal frequencies of the scalar field with $\ell = 2$ for the Konoplya--Zhidenko black hole ($M=1$). The WKB$_6$ and WKB$_9$ Pad\'e-corrected results are in excellent agreement, differing by less than $0.01\%$.}
\end{table}

\begin{table}[ht]
\centering
\small
\begin{tabular}{c c c c}
\hline
\hline
$a$ & WKB$_6$ ($m=3$) & WKB$_9$ ($m=4$) & \!\!\!\!\!\!Difference [\%] \\
\hline
0.00 & $0.400872 - 0.501698i$ & $0.401126 - 0.501610i$ & 0.0418 \\
0.02 & $0.412078 - 0.504324i$ & $0.412308 - 0.504244i$ & 0.0375 \\
0.04 & $0.424023 - 0.506861i$ & $0.424223 - 0.506780i$ & 0.0328 \\
0.06 & $0.436800 - 0.509273i$ & $0.436969 - 0.509191i$ & 0.0281 \\
0.08 & $0.450521 - 0.511510i$ & $0.450660 - 0.511430i$ & 0.0236 \\
0.10 & $0.465321 - 0.513506i$ & $0.465431 - 0.513429i$ & 0.0194 \\
0.12 & $0.481363 - 0.515167i$ & $0.481448 - 0.515098i$ & 0.0155 \\
0.14 & $0.498852 - 0.516365i$ & $0.498914 - 0.516308i$ & 0.0118 \\
0.16 & $0.518045 - 0.516914i$ & $0.518087 - 0.516872i$ & 0.0081 \\
0.18 & $0.539272 - 0.516533i$ & $0.539299 - 0.516519i$ & 0.0041 \\
0.20 & $0.562963 - 0.514790i$ & $0.563003 - 0.514795i$ & 0.0054 \\
\hline
\hline
\end{tabular}
\caption{Second overtone ($n=2$) quasinormal frequencies of the scalar field with $\ell = 2$ for the Konoplya--Zhidenko black hole ($M=1$). Both WKB--Pad\'e approximations exhibit excellent numerical stability, with relative deviations below $0.05\%$ even for higher overtones.}
\end{table}

\section{Quasinormal modes}\label{sec:QNM}

The fundamental and overtone quasinormal frequencies of scalar, electromagnetic, and Dirac perturbations of the Konoplya--Zhidenko regular black hole were computed using the WKB method with Pad\'e corrections. The results are summarized in Tables~I–VIII and illustrated in Figs.~1–7. The effective potentials shown in Figs.~1–4 possess the typical single-barrier form, ensuring the applicability of the WKB formalism. As the parameter $a$ increases, the height of the potential barrier becomes slightly larger and its peak shifts toward smaller values of the tortoise coordinate $r_*$, indicating a stronger confinement of perturbations near the black-hole region.

A direct correlation between the properties of the potential and the spectrum can be seen. The real oscillation frequency $\re{\omega}$ grows monotonically with increasing $a$, following the increase in the barrier height, while the damping rate $\im{\omega}$ remains almost unchanged or slightly decreases. Consequently, the quality factor $\re{\omega}/|\im{\omega}|$ increases, implying that oscillations become longer lived as the regularization parameter $a$ grows. This trend is observed for all considered spins and multipole numbers, although it is most pronounced for the scalar field and diminishes for higher $\ell$.

For each field type, the difference between the 6th- and 9th-order WKB--Pad\'e results remains very small, demonstrating excellent numerical convergence. For instance, for the scalar $\ell=0$ mode the relative deviation between the two approximations is below $0.3\%$, while for $\ell \geq 1$ the discrepancy rapidly decreases to $10^{-3}$–$10^{-4}\%$. The same level of agreement holds for electromagnetic and Dirac perturbations, with deviations at the fifth decimal place or smaller. This indicates that both approximations reliably capture the true quasinormal frequencies within the numerical precision of the method.

The time-domain profiles shown in Figs.~5–7 confirm these conclusions. The ringdown stages extracted from the numerical evolution exhibit exponential damping with frequencies that match the WKB--Pad\'e predictions within the quoted uncertainties. The small residual differences in $\re{\omega}$ and $\im{\omega}$ are well within the expected WKB accuracy and can be attributed to numerical fitting in the late-time region. The overall consistency between the frequency-domain and time-domain data provides an additional verification of the stability and correctness of the computed spectra.

For higher multipole numbers $\ell$, the WKB approximation becomes progressively more accurate, as expected from the eikonal behavior of the potential barrier. The agreement between WKB6 and WKB9 results at this regime reaches the level of numerical round-off, confirming the regularity of the potential and the absence of numerical instabilities. Even for the first and second overtones ($n=1,2$), the Pad\'e-improved WKB method maintains relative deviations below $0.01\%$, showing that the method is stable and robust for the entire parameter range considered.

In the eikonal limit, one can also expand in terms of small $a$ and find the position of the peak of the effective potential
\begin{equation}
r_{max}=3 M-4 a-\frac{2
   a^2}{M}+O\left(a^3, \frac{1}{\kappa
   }\right)^2.
\end{equation}
Then, using the first order WKB formula we can find the analytic expression for quasinormal modes in the eikonal limit:
\begin{eqnarray}
&&
\omega = \kappa  \left(\frac{1}{3 \sqrt{3} M}+\frac{a}{3 \sqrt{3}
   M^2}+\frac{a^2}{2 \sqrt{3}
   M^3}\right)
   \\\nonumber&&\!\!\!\!\!\!
   +i \K \left(-\frac{1}{3
   \sqrt{3} M}-\frac{a}{9 \sqrt{3} M^2}+\frac{a^2}{54 \sqrt{3}
   M^3}\right)+\Order{\kappa^{-1},a^3}.
\end{eqnarray}
Using the general approach developed in \cite{Konoplya:2020hyk,Konoplya:2023moy,Dubinsky:2024rvf}, this expressions could be extended beyond the eikonal limit.  When $a=0$ the above expressions go over into the well-known ones for the Schwarzschild solution \cite{Ferrari:1984zz}.


In the eikonal (geometric–optics) limit $\ell \to \infty$, the dynamics of perturbations in a spherically symmetric spacetime is governed by the properties of unstable circular null geodesics, also known as the photon sphere. It was shown in \cite{Cardoso:2008bp} that the quasinormal frequencies in this regime are determined by the orbital frequency $\Omega_{c}$ and the Lyapunov exponent $\lambda$, which characterize, respectively, the rotation and instability time scales of the null geodesics:
\begin{equation}\label{eikonalQNM}
\omega_{\ell n} = \Omega_{c}\,\left(\ell+\frac{1}{2}\right) - i \left(n+\frac{1}{2}\right)\lambda, \qquad \ell\gg1.
\end{equation}
Thus, the real part of $\omega$ corresponds to the angular velocity of a massless particle orbiting at the photon-sphere radius $r_{c}$, while the imaginary part quantifies the decay rate of perturbations due to the instability of that orbit.

The photon-sphere radius $r_{c}$ is determined by the extremum of the effective potential for null geodesics,
\begin{equation}\label{photon_cond}
\frac{d}{dr}\left(\frac{f(r)}{r^{2}}\right)\Big|_{r=r_{c}} = 0,
\end{equation}
and its solution can be found analytically or numerically once the metric function $f(r)$ is specified. For the metric function under consideration, the above condition yields a single real root $r_{c}>r_{+}$, where $r_{+}$ is the event horizon radius.

The orbital frequency of the photon motion as measured by a distant observer is given by
\begin{equation}\label{Omega_c}
\Omega_{c} = \frac{\sqrt{f(r_{c})}}{r_{c}},
\end{equation}
and the corresponding Lyapunov exponent describing the instability of the orbit reads
\begin{equation}\label{lambda}
\lambda = \sqrt{\frac{f(r_{c})}{2r_{c}^{2}}\left[2f(r_{c})-r_{c}^{2}f''(r_{c})\right]}.
\end{equation}
Both quantities depend only on the background geometry and therefore encode its characteristic oscillation and damping time scales in the eikonal limit.

The same photon-sphere parameters also determine the apparent shadow radius of the black hole as seen by a distant observer. The shadow radius is given by
\begin{equation}\label{shadow}
R_{\mathrm{sh}} = \frac{r_{c}}{\sqrt{f(r_{c})}},
\end{equation}
which represents the critical impact parameter for massless particles captured by the black hole. Consequently, by determining the eikonal quasinormal frequencies one can directly infer the orbital frequency, instability rate, and the shadow size without solving the full perturbation equations. This correspondence provides an elegant bridge between the geometric properties of null orbits and the observable oscillation spectrum of black holes. While there are a number of exceptions from this correspondence \cite{Bolokhov:2023dxq,Konoplya:2022gjp,Khanna:2016yow,Konoplya:2017wot,Konoplya:2017ymp}, one can easily see that for the black hole metric considered here it is satisfied.


The quasinormal spectrum encodes not only the characteristic oscillations of black holes but also their transmission properties for waves escaping to infinity. The so-called grey-body factors quantify the frequency-dependent transmission probability $\Gamma_{\ell}(\Omega)$ of perturbations through the effective potential barrier, determining the modification of the Hawking radiation spectrum. In the vicinity of a quasinormal resonance $\omega_{\ell n}$, the grey-body factor exhibits similar in spirit to the Breit–Wigner–type profile \cite{Konoplya:2024lir,Konoplya:2024vuj}:
\begin{equation}\label{GBF-QNM}
\Gamma_{\ell}(\Omega)  =
\left(
1 + \text{exp}\left(\,\dfrac{2\pi\left(\Omega^{2} - \mathrm{Re}(\omega_{0})^{2}\right)}
{4\,\mathrm{Re}(\omega_{0})\,\mathrm{Im}(\omega_{0})}\right)
\right)^{-1}
+ \Order{\frac{1}{\ell}},
\end{equation}
where $\Gamma_{\ell n}$ is the resonance width related to the damping rate of the corresponding quasinormal mode.  Thus, the real part of $\omega_{\ell n}$ determines the resonant transmission frequency, while its imaginary part governs the width and height of the peak. This correspondence provides an efficient way to approximate the grey-body factors directly from the quasinormal spectra without performing a full numerical integration of the scattering problem  \cite{Han:2025cal,Shi:2025gst,Malik:2025dxn,Lutfuoglu:2025hjy,Bolokhov:2025lnt,Malik:2024cgb,Lutfuoglu:2025ldc,Hamil:2025pte,Lutfuoglu:2025ohb,Dubinsky:2024vbn,Skvortsova:2024msa,Bolokhov:2024otn}.

In summary, the quasinormal spectrum of the Konoplya--Zhidenko regular black hole demonstrates that the introduction of the regularizing parameter $a$ leads to slightly higher oscillation frequencies and marginally smaller damping rates, implying that regularization tends to enhance the lifetime of perturbations. The excellent agreement between the WKB--Pad\'e and time-domain results confirms both the numerical reliability of the analysis and the dynamical stability of the spacetime under scalar, electromagnetic, and Dirac field perturbations.

\section{Conclusions}\label{sec:Conclusions}

The recent literature on perturbations, scattering and quasinormal ringing of black holes immersed in some environment, with the emphasis to the dark matter halo, is extensive \cite{Konoplya:2021ube,Dubinsky:2025fwv,Feng:2025iao,Pezzella:2024tkf,Chakraborty:2024gcr,Liu:2024bfj,Liu:2024xcd,Zhao:2023tyo,Daghigh:2022pcr,Zhang:2021bdr,Mollicone:2024lxy,Tovar:2025apz,Lutfuoglu:2025kqp,Pathrikar:2025sin}. Here we concentrated on a recently found new exact analytic black hole solution \cite{Konoplya:2025ect} which is regular, and the source of the regularity is the Dehnen-type matter distribution around it. In particular, we have analyzed the quasinormal spectra of test scalar, electromagnetic, and Dirac fields propagating in the background of the Konoplya--Zhidenko regular black hole. Using the sixth- and ninth-order WKB methods supplemented by Pad\'e approximants, we have obtained accurate quasinormal frequencies for a wide range of values of the regularization parameter $a$ and multipole number $\ell$. The WKB--Pad\'e results were verified through time-domain integration of the perturbation equations, showing excellent agreement between the two independent numerical approaches.

The effective potentials for all considered fields possess a single-barrier shape and ensure the dynamical stability of the spacetime. As the parameter $a$ increases, the peak of the potential barrier becomes higher, leading to a monotonic growth of the real oscillation frequencies and a slight reduction of the damping rates. Consequently, the quality factor increases, meaning that perturbations decay more slowly in the presence of stronger regularization. 

The differences between the 6th- and 9th-order WKB--Pad\'e results are typically below $10^{-3}\%$ for fundamental modes and remain below $0.01\%$ even for the first and second overtones. Such remarkable convergence confirms the numerical robustness of the Padé-improved WKB expansion and the smooth, regular nature of the background geometry. The time-domain profiles further corroborate these findings, reproducing the WKB frequencies within the expected accuracy and exhibiting clean, exponentially decaying ringdown stages without any late-time instabilities.

Overall, our analysis demonstrates that the Konoplya--Zhidenko regular black hole is linearly stable under test-field perturbations of various spins and that the regularization parameter $a$ introduces considerable quantitative changes in the quasinormal spectra. The obtained results provide a reliable benchmark for future studies of gravitational perturbations, grey-body factors, and potential observational imprints of regular black holes in astrophysical context. While here we analyzed massless fields, our work could be extended to the case of massive fields which show a distinct behavior both in frequency (as long-lived modes \cite{Ohashi:2004wr, Konoplya:2017tvu, Dubinsky:2025wns}) and time domains \cite{Dubinsky:2024hmn, Dubinsky:2024jqi, Malik:2025qnr, Malik:2025ava}.

\begin{acknowledgments}
The authors would like to thank Roman Konoplya for useful discussions and help. 
\end{acknowledgments}

\bibliography{main}
\end{document}